\documentclass[aps]{article}
\usepackage{geometry}                
\usepackage{graphicx}

\usepackage{amssymb}
\usepackage{epstopdf}
\usepackage{amsmath}

\usepackage{amsfonts}
\usepackage{bm}
\usepackage{hyperref}

\DeclareMathAlphabet{\mathpzc}{OT1}{pzc}{m}{it}

\title{Wave-Flow Interactions and Acoustic Streaming}
\author{Clifford Chafin\\\ \small{Department of Physics, North Carolina State University, Raleigh, NC 27695} \thanks{cechafin@ncsu.edu}}

\begin{document}
\maketitle
\begin{abstract}
The interaction of waves and flows is a challenging topic where a complete resolution has been frustrated by the essential nonlinear features in the hydrodynamic case.  Even in the case of EM waves in flowing media, the results are subtle.  For a simple shear flow of constant $n$ fluid, incident radiation is shown to be reflected and refracted in an analogous manner to Snell's law.  However, the beam intensities differ and the system has an asymmetry in that an internal reflection gap opens at steep incident angles nearly oriented with the shear.  For EM waves these effects are generally negligible in real systems but they introduce the topic at a reduced level of complexity of the more interesting acoustic case.  Acoustic streaming is suggested, both from theory and experimental data, to be associated with vorticity generation at the driver itself.  Bounds on the vorticity in bulk and nonlinear effects demonstrate that the bulk sources, even with attenuation, cannot drive such a strong flow.  A review of the velocity scales in the problem suggest that a variation of the Scholte wave at the driver-fluid interface with local cavitation on the decompression phase creates a lateral flow of fluid that generates the stream and imparts the vorticity from the driver plate.  In the case of Darcy flow in sintered media, an analysis of data suggests that acoustically enhanced flow may be the result of surface effects that narrow channels that are cleared by ultrasonic cavitation.  
\end{abstract}

\section{Introduction}
The interactions of waves and flows is complicated by the fact that there are so many kinds of waves and that there is no universality to this problem.  The case of electrodynamics in flowing media is ironically far simpler than the case of hydrodynamic waves.  The interaction of waves in media has been dominated by constitutive law approaches and complicated scattering models that leave it unclear how to proceed.  Even the momentum of an electromagnetic wave in static media has been a long standing problem the resolution of which is still debated.  For condensed matter, propagating EM waves can be considered to be free waves that are partially reflected into a standing wave that generates stress on the media and distinct forces associated with the equilibration at the ends of packets \cite{Chafin-em}.  Accelerating media and deforming media can then be approached by similar methods.  

The case of hydrodynamics is special not just because the waves and flows are composed of the same constituent materials but because hydrodynamics is nonlinear due to the advective nature of the component velocity and density fields.  Whereas the dispersion relations for EM waves in media are well-defined in that standing and traveling waves have the same dispersion relations, the case for acoustic waves has a drift for propagating waves that will be shown to be best understood in introducing an amplitude dependent correction even at lowest orders.  In this sense, hydrodynamic waves have no true linear regime.  The case of surface waves in deep water have a well defined reference frame defined by the deep water velocity but, just as in the acoustic wave case, superpositions are complicated by the drift that increases near the surface.  This helps exemplify that, as Feynman said, when it comes to surface waves, ``everything that can go wrong, does go wrong'' \cite{Feynman}.  

In the current age of metamaterials including photonic crystals and asymmetric materials like light-diodes, it is interesting that the interaction of waves and general flows has not been completely resolved.  Below it will be shown that EM waves in flows can give asymmetric behavior with analogs to Brewster's angle and bound and evanescent bound waves that are only due to flows in otherwise totally isotropic and symmetrical media.  This then leads to an analogous set of behavior for acoustic waves where the effects can be much larger.  Following this, a complete theory of low order interaction of nondissipative acoustic waves with flows is described.  In the course of this, the nature of the nonlinearity is shown to create subtle problems in the decomposition of hydrodynamics into flows and sound.  Once resolved this gives constraints on what sorts of interactions can exist between flow and sound and the kinds of baroclinic sources of vorticity that must be present for acoustic streaming.  The nature of ``Reynolds stresses'' are discussed and shown to be unable, even with attenuation of beam strength, to drive acoustic streaming away from surface boundaries.  

Based on an investigation of Eckart streaming data, theoretical constraints, surface acoustic waves and microfluidic experiments, it is argued that acoustic streaming is the result of Rayleigh acoustic waves and cavitation at the troughs of them that move outwards from the center of the oscillating driver.  This gives a much larger velocity imparted to the fluid that Stokes drift associated with the Airy type waves \cite{Airy} coupled to them at the surface can give.  Similar effects are argued to generate the similarly high velocities seen for droplets driven by surface acoustic waves in microfluidics.  The enhancement of flow from ultrasound for confined networks where Darcy flow dominates is argued to be to an ultrasonic ``cleaning'' effect not acoustic streaming or any forces imparted from the sound itself.

\section{Deforming and Noninertial Dielectrics}
The response of dielectric media to electromagnetic fields is generally considered a rather basic topic in introductory electrodynamics.  Solutions are obtained by boundary and field matching conditions for the macroscopically averaged fields \cite{Jackson}.  Relativistic media are then treated by tensorial treatments and these methods have been extended to give metamaterial responses for cloaking, superlensing and other modern optical devices.  However, the dependence on the classical macroscopic fields has left some holes and enduring confusions in our understanding of this rather basic topic.  The momentum of electromagnetic waves in media has been a matter of dispute since Abraham and Minkowskii in 1905.  Some have even argued that the decomposition of electromagnetic wave and medium response is not uniquely defined \cite{Pfeifer} and that much of the problem is just a matter of consistent boundary conditions.  It is been shown that this is not the case \cite{Mansuripur, Chafin-em} and that such a decomposition, when the wavelengths are large compared to the optical equilibration length of the sample, is unique and provides an exact way to track the flow of energy and momentum exchange between the medium and the fields.  

We are interested in the interactions of waves and flows.  Such a decomposition of electromagnetic waves and medium motion is cleaner in this case compared to the hydrodynamic one so this gives a nice transitional example to the effects of waves on flows before extra complications arise.  As a practical matter, the forces induced by electromagnetic waves on flowing transparent media is miniscule.  The forces exerted by light at intensities that are practical for condensed matter are tiny so we will only be concerned with the lowest order contributions.

The principle descriptor of dielectric response is the dielectric function $\epsilon(\omega)$ which defines the relation $D=\epsilon E$.  This quantity has value through the relation $\nabla\cdot D=\rho_{f}$ to obtain the normal conditions on the electric field.  It can be argued \cite{Chafin-em} that this picture is built on a the notion of a continuum of charge that is displaced and, while convenient for some idealized examples, it introduces confusion when attempting to understand what is actually going on microscopically for wave propagation in media.  Specifically, at an interface of dielectrics, it is better to view the radiators as separated by vacuum and the local polarization as that of polarized dipoles that give no net internal field between them.  Such an approach immediately gives the correct internal stresses of EM waves in a medium and a clear accounting of the momentum sharing between field and medium.  In these gaps the waves have a traveling component with free space relation $c=\omega k$ and a standing wave component that gives the stress and serves as an additional component of stored energy in the medium beyond the energy of the internal oscillators.  The quantity $\epsilon(\omega)$ has a bit of a forced meaning so we utilized the index $n(\omega)$ to give the medium response through $|B_{max}|=\frac{n}{c}|E_{max}|$ for the fields in the gaps between radiators.  
The dielectric constant of uniform isotropic media at rest can be written as 
\begin{align}
n=\sqrt{1+\frac{\rho q^{2}}{\epsilon_{0}(K-\omega^{2}m)}}=\sqrt{1+\frac{\omega_{p}}{(K/m-\omega^{2})}}
\end{align}
where $m$ and $q$ are the mass and the charge of the radiators.  $K$ is the spring constant for each charge.  $\rho$ is the number density of the radiators and $\omega_{p}$ is the plasma frequency.  

 \begin{figure}
  \begin{centering}
 \includegraphics[width=3in,trim=30mm 390 40mm 40mm,clip]{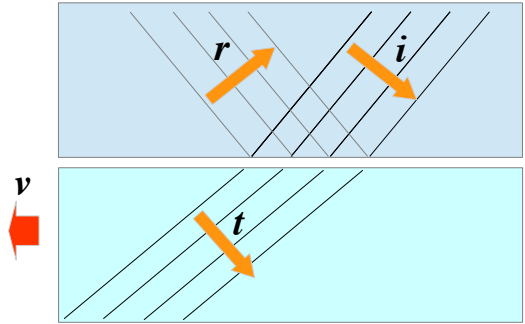}
 \caption{\label{refraction} A pair of shearing transparent blocks of identical dielectric constant with incident, reflected and refracted waves.}
 \end{centering}
 \end{figure} 
 
For this approach to dielectric response to a plane wave at an interface we see that the normal and tangential components of $E$ must be continuous across the boundary.  The magnetic field is discontinuous to the extent dictated by the singular surface currents in $\nabla\times B=\mu_{0}J$ where $J$ is the change in the internal currents at the boundary.  Microscopically such currents exist as well and this is what gives the layer by layer phase shifts that cause the macroscopic wavelength to be shorter than we would expect in vacuum from the relation $c=\omega k$, however, these give an incremental variation when averaged over the sample except at such a boundary as where there is a discontinuous change in the dielectric.

\subsection{Shear Flow and EM Waves}\label{em}
Let us now consider two blocks on equal dielectric index and separated by a thin ($<\lambda$) layer of index matched fluid.  The lower block now slides to the left at velocity $v$.  An incident wave moves in at angle $\theta$ (from the surface not the normal\footnote{We have chosen this instead of the usual angle from the normal since will later be interested in cases where the flow shear and waves are nearly parallel.}) and we presume it reflects and refracts as in fig.~\ref{refraction}.  The lower block is an example of a moving dielectric where we are viewing the fields in rest frame which is not commonly seen.  If we were to boost it to the rest frame the fields would be reoriented.  Note that these are the microscopic $E$ and $B$ fields so they transform by the usual Lorentz boost equations ,however, in a boosted frame, the dielectric response becomes tensorial and the density of oscillators changes so such a transformation must be done with care.  The dielectric response changes due to the denser number of radiators and an increased (and anisotropic) stiffness in the restoring charges of the block in the boosted frame.  This will automatically lead to a reflected wave.  However, these effects can be compensated for by using a less dense material with more weakly bound charges but, nevertheless, there will still be a reflected wave and altered angle of transmission due purely to the retarded effects and the interaction of the wave with the flow.  

 \begin{figure}
  \begin{centering}
 \includegraphics[width=3in,trim=20mm 340 10mm 20mm,clip]{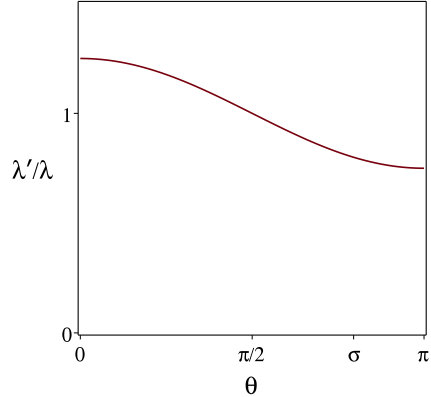}
 \caption{\label{theta} The ratio of wavelengths of the transmitted and incident waves.  Thes line corresponds to $\alpha=0.25$.}
 \end{centering}
 \end{figure} 
  \begin{figure}
  \begin{centering}
 \includegraphics[width=3in,trim=20mm 340 10mm 20mm,clip]{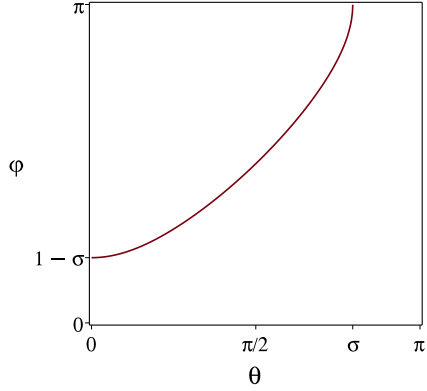}
 \caption{\label{phi1} Allowed angles of refraction versus incident angle of radiation.  Note these are relative to the surface not the normals. }
 \end{centering}
 \end{figure} 
We can match the velocity of the waves at the shear interface and the wavelengths to give:
\begin{align}\label{match}
\frac{c}{n}\frac{1}{\cos(\theta)}&=\frac{c}{n}\frac{1}{\cos(\varphi)}-v\\
\lambda\cos(\theta)&=\lambda'\cos(\varphi).
\end{align}
This gives a relation of the change in wavelength of the waves as
\begin{align}\label{match1}
\frac{\lambda'}{\lambda}=1+\frac{v n}{c}\cos(\theta)
\end{align}
as is illustrated in {fig.\ \ref{theta}}.  Interestingly we see a gap in the range $\theta\in I$  where 
\begin{align}\label{gap1}
I=[\sigma,\pi]\\
\frac{1}{\cos(\sigma)}+\frac{v n}{c}=-1
\end{align}
 \begin{figure}
  \begin{centering}
 \includegraphics[width=2in,trim=30mm 300 20mm 50mm,clip]{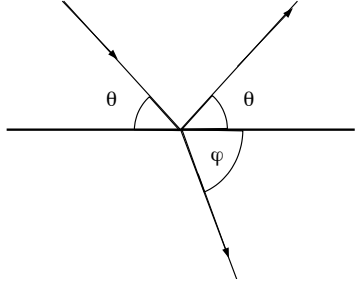}
 \caption{\label{angle} Angles of the incident, transmitted and reflected waves.}
 \end{centering}
 \end{figure} 
where there must be total internal reflection.  This follows from the solution for the refracted angle 
\begin{align}\label{varphi}
\varphi=\cos^{-1}\left(\frac{1}{\frac{1}{\cos(\theta)}+\frac{v n}{c}}\right)
\end{align}
as illustrated in fig.~\ref{phi1}.  The gap in transmission given by $I=[\sigma,\pi]$ gives total internal reflection but, unlike a dielectric difference, gives an asymmetric response.  The parameter that controls the quantitative shape of these functions is $\alpha=\frac{v n}{c}$.  Since we are interested in nonrelativistic effects but still ones that are large enough to illustrate graphically, a value of $\alpha=0.25$ has been used.  

Interestingly we can make an analog to Snell's law and give an effective (angle dependent) index of refraction as follows.  The usual form of Snell's law in terms of surface angles (in place of normal angles) is
\begin{align}
\frac{\cos(\theta_{t})}{\cos(\theta_{i})}=\frac{n_{i}}{n_{t}}
\end{align}
so that we can define a fictitious index of 
\begin{align}
n_{\text{eff}}=n(1+\alpha\cos(\theta)).
\end{align}
Although this gives the correct angles for the wavefronts, it does not give the gaps in the transmission angles and, as we will now see, gives incorrect field amplitudes that are transmitted.  

The field intensities can be found in the p-polarization (TM wave) case by matching the electric field intensities of the incident, reflected and transmitted waves.  Using the coordinates as in fig.~\ref{angle} we have
\begin{align}\label{E}
\bar E_{i}&=E_{i}(\sin(\theta),\cos(\theta))\\
\bar E_{r}&=E_{r}(-\sin(\theta),\cos(\theta))\\
\bar E_{t}&=E_{t}(\sin(\varphi),\cos(\varphi))
\end{align}
and the $B$ fields are all perpendicular to the plane so 
\begin{align}
B_{i}&=\frac{n}{c}E_{i}\hat z\\
B_{r}&=\frac{n}{c}E_{r}\hat z\\
B_{t}&=\frac{n}{c}E_{t} \hat z
\end{align}
\begin{figure}
  \begin{centering}
 \includegraphics[width=3in,trim=20mm 340 10mm 20mm,clip]{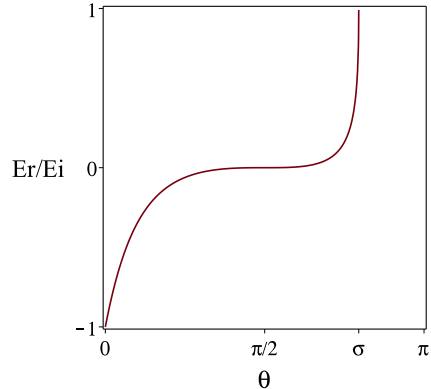}
 \caption{\label{Er} The intensity of the electric field of the reflected wave relative to the incident one.}
 \end{centering}
 \end{figure} 
The boundary conditions $E_{||}$ and $B_{||}$ are continuous at the surface give
\begin{align}
(E_{i}-E_{r})\sin(\theta)&=-E_{t}\sin(\varphi)\\
E_{i}+E_{r}&=-E_{t}\\
E_{i}(\sin(\theta)-\sin(\varphi))&=E_{r}(\sin(\theta)+\sin(\varphi))
\end{align}
hence the transmitted intensity is related to the incident intensity by 
\begin{align}
E_{r}=E_{i}\frac{\sin(\theta)-\sin(\varphi)}{\sin(\theta)+\sin(\varphi)}
\end{align}
as illustrated in fig.~\ref{Er}.  
The surface charge density is given by the discontinuity in the normal electric field as $\Delta E_{\perp}=\epsilon^{0}\rho_{s}$ so that $\rho_{s}(x,t)=\epsilon_{0}^{-1}(E_{i}+E_{r})(\cos(\theta)-\cos(\phi))\sin(k'x-\omega t)$ where $k'=k\cos(\theta)$ is the projected wavevector on the x-axis.

\section{Sound and Flows}
The previous example was meant as an introduction to the effect of shear flow on EM waves to demonstrate some of the complexity that can arise even in the case where there is no important advective term as there is the case of sound waves.  Of course, realistic shear flows for this case are extraordinarily small.  We can give a measure of this based on $\lambda |\nabla v|/c$ where $\lambda$ is the wavelength of light we are considering.  The refractory band gap, $I$, is then very small however, its existence does suggest that shear flows parallel to waves can exert a collimating effect.  Such an effect would potentially be of interest in the case of acoustic streaming and in the case of ultrasound enhanced flow in narrow channels where Darcy flow can locally generate strong shears.  

Let us first consider the case of the interaction of sound and flows in the the case of no dissipation.  Given that the relative rates of sound waves and flows are generally separated by many orders of magnitude, this is a challenging subject to study experimentally.  Even the case of surface waves interacting with flows has many unanswered questions where the time scales are much closer and longer.  One confounding feature is that the solutions of sound in terms of periodic eulerian pressure and velocity solutions give an amplitude dependent lagrangian mass flux.  This means that, at the level where we are concerned about momentum and mass conservation, the system is intrinsically nonlinear.  We can send sound through a closed chamber with fixed boundaries and can turn these on and off very rapidly.  Acoustic streaming invariably introduces vorticity and these flows build up gradually, often on the order of minutes.  Vorticity can only be introduced in the bulk by baroclinic means or at the boundaries and acoustic streaming experiments seem to demonstrate that the vorticity initially builds up at the oscillator itself but the reason for this has been elusive and the magnitude of the effect is strangely large.  A similarly large effect has been observed for surface acoustic waves driving droplets and microfluidic flows \cite{Yeo}.  

For a liquid in a chamber initially at rest where a traveling sound beam or pulse is abruptly turned on there is no immediate creation of vorticity at the edges of the beam and no net flow generated.  To use the traveling wave basis of solutions to build the beam we then must compensate with a $w(x)=-v_{d}(x)$ compensating field that cancels the net drift locally at each point.  What is most strange is that this local drift evolves purely in response to the often rapid changes of the sound and not the usual hydrodynamic equations.  For this reason we might best call it ``faux-drift'' since a packet of sound has no such drift and the compensation must be present in an exact hydrodynamic evolution but that we have to insert ``by hand'' to ensure this even at the expense of such strange evolution.  We might best think of this faux-drift as part of the redefinition of the background flow that is necessary to give a consistent finite order approximation in terms of a sound-flow decomposition while keeping the dispersion relation that holds for standing waves (where amplitude dependent corrections don't arise at this order).  

Now let us impose an initial (inviscid) flow $u(x)$ that is slowly changing and may contain vorticity.  As a first case, let it be a completely rotational field so that a Helmholtz decomposition of it gives no irrotational part.  This flow evolves according to Navier-Stokes (N-S) in a manner independent of the sound field.  The vorticity moves according to the vorticity transport equation but the sound and $w(x)$ fields cancel and there is no baroclinic source so the flow is unchanged by the sound.  The sound, on the other hand, is altered by the flow in a manner that cannot exchange energy or momentum between them.  For liquids that have no density variability on time scales longer than the sound variations, this allows the flow deformations to adiabatically reshape and translate the density variations and velocity fields (local oscillating velocity + faux-drift $w(x)$).  The kinetic energy of the sound component and potential energy of the pressure fields are then unchanged but get dragged to a new frequency distribution.  The rate of this deformation is given by the ratio of the local flow gradient over a wavelength $\lambda |\nabla v|$ relative to the period $\tau$.  This decomposition is made possible by this very large separation of these scales but we accept that, while some small acoustic effects can give cumulatively growing results, these do not do so for an exchange of flow and acoustic energy in this case by the above argument.  Note that the compensating flow $w(x)$ evolves to keep the acoustic motion stationary relative to the flow $u(x)$ and so the net acoustic field has no momentum \cite{McIntyre, Chafin-waves}.  

\subsection{Sound in Flows Without Dissipation}\label{drift}

It is frequently convenient to evolve sound waves in terms of a basis of oscillatory waves since we expect the coefficients of expansion to evolve slowly due to nonlinear and other effects.  Let us consider acoustics at the level of sinusoidal expansions on an external flow $V(x)$.  Since this can include translations and rotations that lead to motions that are fast relative to the wave speed in the rest frame of the medium, it is advisable to pull these out and investigate the system in such a frame.  If this allows the density of the medium to be nearly constant when averaged over times scales longer than the acoustic variations and the remainder of the flow velocity, $u(x)$, is small over the relevant volume of interest we can consider a convenient expansion as follows.  The basis of velocity potential in these coordinates is given by 
\begin{align}
\phi_{k}(x,t)&=\sin( k\cdot(x-v_{k}t))\\
\tilde\phi_{k}(x,t)&=\cos( k\cdot(x-v_{k}t))
\end{align}
where, for most purposes the dispersion relation $|v_{k}|=c_{s}$ is adequate.  The compensating drift velocity of a flow at a point $x$ is given by the net lagrangian flow associated with $v(x,t)=\sum_{k}a_{k}\nabla \phi_{k}(x,t)+b_{k}\nabla \tilde\phi_{k}(x,t)$.  This flow, when smoothed by filtering over a number of oscillations, is 
\begin{align}
\langle w(x,t)\rangle=-\langle\sum_{k}\left(a_{k}^{2}k^{2}v_{k} \cos^{2}(k\cdot(x-v_{k} t))+b_{k}^{2}k^{2}v_{k} \sin^{2}(k\cdot(x-v_{k} t)) \right)\rangle
\end{align}
where the brackets imply averaging over the length scale of typical oscillations.  The remaining flow must be included in the infinitesimal evolution equations as well.  For an infinitesimal evolution of $\tilde\phi_{k}$ we have 
\begin{align}
\tilde\phi_{k}(x,dt)&=\cos(k\cdot(x-v_{k} dt +w(x)dt-u(x)dt))\\
&\approx\cos(k \cdot x)-\sin(k \cdot x)k\cdot(v_{k}+w(x)-u(x)) dt
\end{align}
 and similarly for $\phi_{k}(x,dt)$.  Ideally, we would like a fixed spatial basis and transfer all time evolution to the coefficients of expansion.  Let the spatial basis functions be
\begin{align}
\varphi_{k}(x)&=\sin( k\cdot x)\\
\tilde\varphi_{k}(x)&=\cos( k\cdot x)
\end{align}
and let the general potential fields be given by $\phi(x,t)=\sum_{k}a_{k}(t)\varphi_{k}(x)$.  The evolution equations for the free linear case is
\begin{align}
\dot a_{k}&= + (k\cdot v_{k}) b_{k}=+\omega_{k} b_{k} \\
\dot b_{k}&=- (k\cdot v_{k}) a_{k} =-\omega_{k} a_{k}.  
\end{align}
To incorporate the compensating faux-drift and external flow we should encode these in terms of the basis.  Define the functions 
\begin{align}
u_{kl}&=\langle u(x)\cdot k | \varphi_{l}(x) \rangle \\
\tilde u_{kl}&=\langle u(x)\cdot k | \tilde\varphi_{l}(x) \rangle 
\end{align}
as the projections of the $k$ components of $u(x)$ on the $\varphi, \tilde\varphi$ basis.  Since $u(x)$ changes slowly or may be constant for many problems, this computation may only have to be done rarely or once.  In contrast, the faux-drift can change as rapidly as the mean acoustic field can be altered so will have to be computed and evolved in concert with it.  The projections of $k\cdot w(x)$ on the basis gives
\begin{align}
w_{kl}&=\langle w(x)\cdot k | \varphi_{l}(x) \rangle= 0\\
\tilde w_{kl}&=\langle w(x)\cdot k | \tilde\varphi_{l}(x) \rangle=-\frac{1}{2}\langle\left(a_{k}^{2}k^{2}\omega_{k} (\tilde\varphi_{0}+\tilde\varphi_{2k}) +b_{k}^{2}k^{2}\omega_{k} (\tilde\varphi_{0}-\tilde\varphi_{2k})\right)|\tilde\varphi_{l}\rangle\\
&=
-\frac{1}{2}k^{2}\omega_{k}\left(a_{k}^{2}  -b_{k}^{2}\right)\delta_{2k,l}
\end{align}

sing these we have the low order local drift corrected equations of motions
\begin{align}
\dot a_{k}&=+\omega_{k} b_{k}+\sum_{l}\tilde w_{k,l}-\sum_{s}\tilde u_{k,s}=+\omega_{k} b_{k}+\tilde w_{k,2k}-\sum_{s}\tilde u_{k,s} \\
\dot b_{k}&=-\omega_{k} a_{k}-\sum_{l}w_{k,l}+\sum_{s}u_{k,s}=-\omega_{k} a_{k}+\sum_{s}u_{k,s}.  
\end{align}

The equations never generate any vorticity to this order.  Later we will see how pressure corrections from nonlinear terms can generate vorticity but the forces that drive this are external pressures that are not associated with acoustics.  
These observations tell us that ultimately any explanation of acoustic streaming must be baroclinic in nature and found on the scale of the sonic wavelengths rather than in an interaction among waves and flows.

\subsection{Shear Flow and Sound}
Let us now reconsider the examples of sec.~\ref{em} with a shear flow but now with an acoustic field instead of an electromagnetic wave.  The velocity field is given by a potential $v=\nabla\varphi$ on each side of the shear layer.  The wavelength matching conditions, eqns.~\ref{match}, \ref{match1}, normal gap eqn.~\ref{gap1}, and refraction angle eqn.~\ref{varphi},  are the same assuming $n=1$ and $c\rightarrow c_{s}$.  This gives the analogous relations
\begin{align}\
\frac{c_{s}}{\cos(\theta)}&=\frac{c_{s}}{\cos(\varphi)}-v\\
\lambda\cos(\theta)&=\lambda'\cos(\varphi).
\end{align}
and wavelength matching relation
\begin{align}
\frac{\lambda'}{\lambda}=1+\frac{v}{c_{s}}\cos(\theta).
\end{align}
so that these relations are the same with a new definition of $\alpha=v/c_{s}$.  

\begin{figure}  
  \begin{centering}
 \includegraphics[width=3in,trim=20mm 340 10mm 20mm,clip]{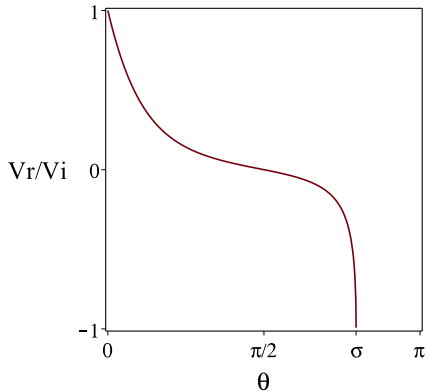}
 \caption{\label{vr} The magnitude of the velocity of the reflected wave relative to the incident one for the case $v/c_{s}=0.25$.}
 \end{centering}
 \end{figure}

The matching conditions for the velocity fields now give
\begin{align}
\bar v_{i}&=v_{i}(-\cos(\theta),\sin(\theta))\\
\bar v_{r}&=v_{r}(\cos(\theta),\sin(\theta))\\
\bar v_{t}&=v_{t}(-\cos(\varphi),\sin(\varphi)).
\end{align}
(The pressure fields follow by constraint: $\nabla^{2}P=-\nabla\cdot(\rho \dot v)$.)  In the acoustic case, we need both $v_{||}$ and $v_{\perp}$ continuous so 
\begin{align}
(v_{i}-v_{r})\sin(\theta)&=v_{t}\sin(\varphi)\\
(v_{i}+v_{r})\cos(\theta)&=v_{t}\cos(\varphi)\\
(v_{i}-v_{r})&=(v_{i}+v_{r})\cot(\varphi)\tan(\theta)
\end{align}
\begin{align}
v_{r}=v_{i}\frac{\sin(\phi)\cos(\theta)-\cos(\phi)\sin(\theta)}{\sin(\phi)\cos(\theta)+\cos(\phi)\sin(\theta)}
\end{align}
The relative size of the velocity of the refracted wave to the incident one is plotted in fig.~\ref{vr}.  We see that this is qualitatively the same as the electromagnetic case but less steep.  

\section{Acoustic Streaming}
There are two red herrings in the subject of acoustic streaming.  The first is the notion that sound carries momentum and therefore generates a mass flux.  Although this has been refuted above as a artifice of a problematic decomposition subtlety of wave and bulk translation, it should be mentioned that there is an analogous effect in the subject of surface waves.  Stokes drift is present in all irrotational solutions of advancing wave motion.  There is an analogous complication that, in all wavetank and oceanic experiments, no net drift of wave motion is detected.  This is still an unresolved issue \cite{Monismith} but the author has recently argued it is a subtlety of transients at the leading edges of packets and wave trains that lead to microbreaking in a fashion that must completely balance the net mass flux between the waves and backwards surface shear diffusing into the depth \cite{Chafin-waves, Chafin-windwaves}.  The second distraction is in the notion of ``wave stress'' whereby sound has some hidden stress from nonlinear terms.  In the inviscid limit, there is only pressure.  Reynolds stresses are a distraction of averaging that cannot give any net motion in this fashion \cite{Chafin-acoustic}.  Even such nonlinearly induced pressures can be easily cancelled by elastic expansion of the nearly incompressible fluid carrying the waves.  It is neglecting this long range reequilibration that has led to the notion of Reynold's stress driving flows.  We will show that pressure gradients that drive Poiseuille flow in a tube and nonlinearly induced density gradients give baroclinic sources in the bulk of the flow that give forces that speed up the flow in a tube in the region complementary to the sound beam.  However, in the absence of such a driving pressure, there is no such baroclinicity hence it cannot be what drives acoustic streaming.  

A reasonable litmus test for having found the source of acoustic streaming is that there must be a suitable baroclinic source for the induced vorticity.  In addition to this, I propose the following hydrodynamic principle.  For a nonrotating hydrodynamic system, the mean velocity fields in a quasistationary flow driven by motion at the boundaries will never exceed the velocities of the boundaries themselves.  Certainly, we can focus a sound beam and create locally faster velocities in the interior but these are not the mean velocities that average out the sound oscillations.  The only way fluids can be accelerated beyond this speed is through a pressure gradient that will then be balanced in becoming stationary.  Although I will not be presenting a proof of this, the following examples will lend support to such a notion.  Several important examples have appeared that make an understanding of vorticity and surface interactions crucial.  The usual quartz wind experiments arise where the flows can be far faster than the drift rates at the walls due to Rayleigh surface acoustic waves interfacing with hydroynamic ones.  These are often called Scholte waves.  The motion of small droplets and microfluidic flows are other examples \cite{Moudjed, Yeo, Brennen}.  In the case of motion in small channels, the evidence for ``exclusion zone'' (EZ) water is now conclusive \cite{Pollack}.  These layers can extend up to 0.1mm near some polymer surfaces.  This fourth phase of water has been shown to have a density of 10\% more than ordinary water, an increasing rigidity that might be characterized as viscosity and an enhanced exclusion of bacteria and colloids in a manner reminiscent of wall exclusion used for sorting and the Zweifach-Fung effect \cite{Zweifach, Fung}.  

\subsection{Limits on Baroclinic Sources}
The notion of Reynold's stresses has played the role of a panacea for hydrodynamicists when mysterious large forces or complications of turbulence arises.  It is most commonly used in closure schemes for turbulent flows where some time averaging is preserved.  The warning that should accompany it, is that they are often used when there is small viscosity and the actual shears are therefore small so the physics they encode is not much associated with shear forces.  In the real time turbulent flow the circulating motion drives expanding and stretching flows that conserve vorticity in this case.  Oscillations of these motions drive oscillating pressure fields which are determined in a very nonlocal fashion for incompressible fluids.  The local time averaged shear quantity has to adequately incorporate these long range effects in a local quantity for the mean flow.  How to control and place limitations on its use is still not understood.  Most faith in it comes from the long precedent of its use and the lack of many other means to do  computations in this area.  In the case of streaming let us analyze the case of a beam in a standing tube of water and determine the effect of the nonlinear terms associated with Reynolds ``stress'' and examine its effect.  

\begin{figure}
  \begin{centering}
 \includegraphics[width=3in,trim=20mm 340 10mm 20mm,clip]{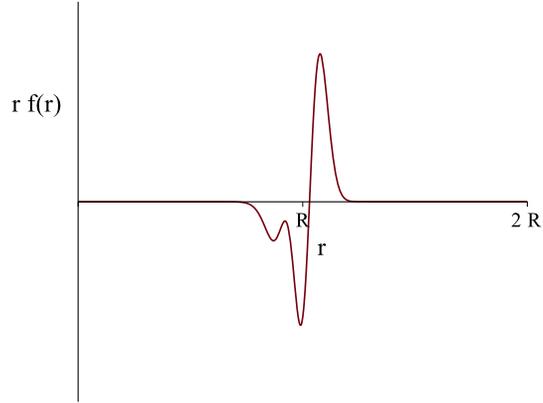}
 \caption{\label{rf} The value of the nonlinear source $r f(r)$ for the nonlinear pressure term. }
 \end{centering}
 \end{figure}

\begin{figure}
  \begin{centering}
 \includegraphics[width=3in,trim=20mm 340 10mm 20mm,clip]{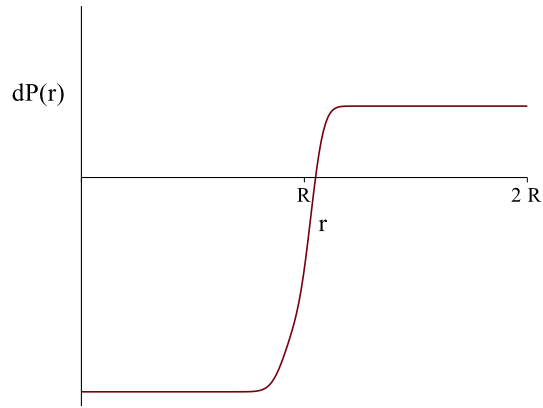}
 \caption{\label{P} Pressure correction from nonlinear contributions.}
 \end{centering}
 \end{figure}

Consider a cylindrical cavity or radius $R_{0}$ filled with liquid at pressure $P_{0}$ with a circular radiator of radius $R$ at one end and perfect absorption at the other.  The wavelength of the sound is much smaller than the cavity length $L$ and the radius $R_{0}$ so that spreading can be neglected.  Note that the cavity can be chosen very short so this condition holds very well.  This does not imply that we can ignore lateral motions at the edge of the beam as we will see.  The pressure of the fluid, on short scales is given by the local pressure oscillations of the sound beam.  On the larger scale we can compute them from the time average of the nonlinear correction to the pressure.  To second order we can compute this long range pressure $\bar P$ from the divergence of the N-S equations
\begin{align}
\nabla^{2}\bar P=-\rho\langle \nabla\cdot(v\cdot\nabla v)\rangle=-\rho f(r).
\end{align}
We choose a smooth cutoff for the lateral motion of size $R$ so the velocity potential of the sound field is given by $\phi=a \sin(k z-\omega t) [1+\text{erf}(1+\mbox{\small\( \frac{r^{2}}{R^{2}}\)})]$.  In cylindrical coordinates we have
\begin{align}
\nabla^{2}\bar P=\rho a^{2}\frac{1}{4\pi^{3/2}r R^{6}} \bigg(
 \pi R^{3}k^{2}[(R+r)^{2}-3r^{2}](1+\text{erf}(-1+\mbox{\small\( \frac{r^{2}}{R^{2}}\)}))e^{-1+\frac{r^{2}}{R^{2}}}\\
    +  (8(r-R)^{2}r+  R^{2}[(R^{2}k^{2}-4)r+2R])e^{-2(1-\frac{r^{2}}{R^{2}})}   \bigg)e^{\frac{2r(R-r)}{R^{2}}}.
\end{align}
The solutions for $rf(r)$ and $\bar P(r)=\delta P$ are given in fig.~\ref{rf} and fig.~\ref{P}.  
These show that radial corrections are induced in the pressure field that seem to give a reduction in the pressure on the interior of the beam.  Should there be an attenuation of the beam with distance we might interpret this as a source of force along the length of the beam.  Of course such a pressure variation is not sustainable and must be immediately matched by a radial and longitudinal shift in mass that leads to an increase in density in the beam support at the expense of the complement.  The extent to which this occurs depends on the compressibility of the fluid and the width of the cavity.  Since the compressibilities of liquids are very small, this is generally a tiny effect.  At the interface of two fluids where a beam passes from one to another there are some interesting variations in elevation over the width of the beam that can be upwards or downwards depending on the various sounds speeds in the liquids \cite{Beyer}.  It should be remembered that $\rho$ was the density of the fluid when it was constant and before the acoustic beam was present.  

There is a structure to the nonlinear pressure source term $f(r)$ that is not immediately obvious.  The integral $\int r f(r)dr=0$ so that this acts as a kind of ``double layer'' of pressure ``charge.''  Examination of the cylindrical form of the equations gives\footnote{The contributions from $\nabla\rho$ are higher order and ignored here.} 
\begin{align}
U=v\cdot\nabla v=\left( v_{r} \frac{\partial v_{r}}{\partial r}+v_{z} \frac{\partial v_{r}}{\partial z}\right)\hat r+\left( v_{r} \frac{\partial v_{z}}{\partial r}+v_{z} \frac{\partial v_{z}}{\partial z}\right)
\end{align}
and 
\begin{align}
\nabla\cdot U=\nabla\cdot (v\cdot\nabla v)=\frac{1}{r}\frac{\partial}{\partial r}(r U_{r})+\frac{\partial}{\partial z}U_{z}
\end{align}
so that $\nabla^{2}P=-\rho \nabla\cdot U$.  Only the radial terms contribute and the first gives a double layer term.  The second gives a triple layer so that it only contributes a vertical shift overall.  This first term for sharp beam edges gives a double layer that is well approximated by a pair of delta functions
\begin{align}
\nabla^{2}P=-\rho\frac{a^{2}}{R}\bigg(\delta(r-R-\epsilon)-\delta(r-R+\epsilon) \bigg)
\end{align}
so that the pressure correction is given by 
\begin{align}\label{dP}
\bar P=\rho \frac{a^{2}}{8R} \Theta(r-R)+C
\end{align}
where $\Theta$ is the Heaviside step function and $C$ is a constant to give mass conservation.  This then induces the elastic relaxation to increase density inside the beam.  When an external pressure field, as drives Poiseuille flow, is present, this leads to a baroclinic vector that slows the interior of the flow relative to the exterior and alters the usual parabolic profile.  This is generally a very small density gradient so has limited ability to modify the flow much.  There is furthermore no evidence that such an effect can produce the kinds of flow we observe in Eckart streaming or the kinds of enhancement of flow we see from ultrasonic stimulation of fluid in sintered materials.  Most importantly, by examining the so-called Reynolds stresses from this point of view of small scale averaging, we see that they are always cancelled by small elastic flows and so are not ``stresses'' at all but simply generate long range corrections to the density.  

As a final comment, note that when $R_{0}\gg R$ the density correction primarily increases the density of the beam support since the mass of the complement is so much larger.  In the opposite case where $R_{0}$ is only slightly larger than the beam support, $R_{0}=R+\epsilon$, the decrease in density in the complement may be large enough to create cavitation unless the background pressure is large enough.  These lateral variations in density are sinks on the energy of the beam as it ``rings-up'' to its full intensity from transients.  Such additional energy storage is important for packet motion, dissipation and can play a role in the group velocity of packets \cite{Chafin-em} and give another avenue for nonlinear effects to become large.  Since these effects are determined by the geometry in the complement of the beam they are not derivable from dispersion relations or any local analysis.

\subsection{Eckart Streaming}
In the case of acoustic streaming in an open fluid the transients (\hyperlink{http://www.lmfa.ec-lyon.fr/perso/Valery.Botton/english/videos_streaming.html}{link}) reveal that the flow begins near the oscillator and grows outwards as a narrowly focused jet from it.  The fluid is drawn in from the sides and moves out as a jet that generates tightly confined back flow around it that are not well associated with the support of the beam itself.  This suggests that the source of the vorticity is entirely at the driver.  Let us now look at the nature of the interface between the oscillator and the fluid.  The driver is excited at the center so we expect some outwards moving Rayleigh surface waves on it.  The presence of the fluid adds extra mass that lowers the propagation rate.  Much is often made of the difference in sound speeds between media in generating forces but the interface has pressure move across it and the motions become a single correlated wave.  The Scholte waves are the waves describing Rayleigh surface acoustic waves interfacing with hydrodynamic Airy waves that then give a modified dispersion relation.  

\begin{figure}
  \begin{centering}
 \includegraphics[width=3in,trim=20mm 220mm 20mm 20mm,clip]{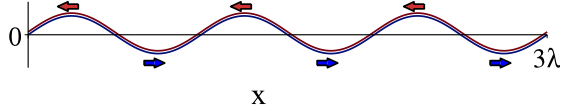}
 \caption{\label{interface} An Airy fluid wave surface (blue) and solid surface Rayleigh wave (red) matched to give a Scholte wave.  The net waveform is moving left.  }
 \end{centering}
 \end{figure}

As we discussed before, the distinction between standing and traveling hydrodynamic waves is important and these should not be thought of as having exactly the same dispersion relations even at very small amplitudes since the traveling waves have amplitude dependent propagation that must be corrected for.  In the case of interfacial waves between liquids or liquid and solid, there is also the problem of vorticity generation and damping at the interface.  For interfacial waves in a stratified liquid, this occurs for both standing and traveling waves since they must have opposite motions at the interface.  A smaller effect is that the peak of the crest must move in the direction of a propagating wave and this is the effective trough of the wave of the fluid above it.  For the case of a solid and liquid interface, as in fig.~\ref{interface}, the Rayleigh wave has a motion opposite that of the wave motion at the crest.  This allows an Airy and Rayleigh wave to interface and both move in the same direction.   
For standing waves this works perfectly but for traveling waves there is the problem of the Stokes drift of the wave.  Unlike the acoustic version, this drift is not an artifice of some decomposition of waves and flows that can be removed.  The far from surface region gives a well defined notion of ``at rest'' for the fluid.  This drift is real and, since the solid cannot match it, gives a source of drag at the interface that is supplementary to the viscous losses that also must manifest at the surface of Airy waves.  

For MHz frequencies, typical wavelengths are in millimeters.  The amplitude of waves are given by their intensity 
\begin{align}
I=\rho v^{2} c_{s}
\end{align}
where the speed of sound is $c_{s}\sim 10^{3}$m/s and $\rho\sim 10^{3}$kg/m$^{3}$ in most relevant condensed matter.  The intensity of many such experiment are of the order of W/cm$^{2}$.  This gives a particle surface velocity of $a\omega=v\sim 10^{-1}$m/s where, here, $a$ is the size of the particle displacement (not the magnitude of the velocity potential as before).  However, this is an oscillatory velocity and not the drift velocity of fluid that can get dragged with the wave.  The Stokes drift is reduced by a factor of $a/\lambda=10^{-4}$.  However, we frequently get flows on the order of 10's of cm/s.  Since there is no pressure field or stress to drive them up to such speeds we are faced with the only velocity comparable to the flow speed being the oscillating surface speed.  

\begin{figure}
  \begin{centering}
 \includegraphics[width=3in,trim=20mm 200mm 20mm 20mm,clip]{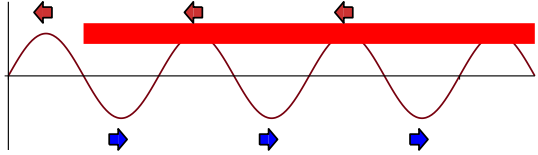}
 \caption{\label{slab} The trough-cavitation crest conveyor: A rigid floating slab riding the crests of an Airy wave or a Rayleigh wave.  In the former case it travels as the particle crest velocity in the wave direction.  In the later, it moves against the wave motion.  }
 \end{centering}
 \end{figure}

The Stokes drift of surface waves is comparable to the velocity of the particles, $a\omega$, but reduced by an order of wave slope $a/\lambda$.  This means that floating matter will get moved by waves at a much slower rate than the particles are moving in the wave.  However, if we consider a floating rigid mat that spans several crests, as in fig.~\ref{slab}, the velocity the mat experiences is now the surface velocity of the crests $a\omega$.  A similar effect occurs for Rayleigh waves where a similar body on the surface now feels the crest velocity.  However, now the motion is opposite to the direction of advance of the waves.  These provide a mechanism for velocity of particle motion to be much larger than averaging would predict.  

It has been shown that the vibration of a bubble at a solid liquid interface increases the speed of heat flux across the boundary \cite{Gould}.  This is thought to be the result of acoustic streaming but the length of the flows in such a case are on the scale of the bubble size so it is not clear if an increase in convection can really be responsible.  One possible explanation is that the the EZ layer is disrupted and this affects the surface convection.  Another is that the pressure drop enhances cavitation in the vicinity and improves thermal coupling at the interface by interfering with the no-slip conditions there.  Let us test the viability of such a model.  The pressure from the sound can be found by the relation $I=P^{2}/\rho c_{s}$ so $P\sim 10^{5}$Pa.  This is comparable to atmospheric pressure so that, given sources of nucleation, such as might be found at an interface, cavitation is possible.  There are two confounding features here.  The plate is driving sound perpendicular to it and Rayleigh waves are moving outwards from the center.  It is not clear what the intensity of the Rayleigh waves are.  We have simply used the estimates from the longitudinal waves to get an order of magnitude.  Such a situation would create cavitation at the troughs of the sound waves so that contact is now at the crests and violate no-slip boundary conditions.  This would create a surface drag at the characteristic flow speed of $10^{-1}$m/s.  One test for such a mechanism is to apply a larger external pressure on the system so that cavitation is reduced and see if it reduces the incidence of streaming.

There is some evidence for such a dynamic in the case of droplets driven across surfaces at 1-10 cm/s due to nanometer sized oscillations of Rayleigh waves in the 1-10 MHz regime \cite{Yeo}.  These also cannot possibly give drift velocities that match these speeds so it seems that some direction biased impulse from the surface motions are being transferred to the fluid.  In this case the droplets are moving in the direction of the motion of the waves so, by the above mechanism, would have to be getting preferential contact with the troughs.  Explanations for these motions are often phrased in the language of impulse transferred by the sound even though some of these same articles emphasize the Westervelt paradox to emphasize that the eulerian and lagrangian velocities are not the same so that sound need carry no momentum.  This seems to be the paradox of the use of Westervelt's paradox.  Apfel and Chu \cite{Apfel} gave an explanation of this paradox in terms of a density imbalance of the sound wave moving forwards and returning.  However, if one is to track every particle in lagrangian fashion, this is problematic.  The Stokes drift in the surface wave component would drag fluid backwards against the surface acoustic wave and there is a circulating flow seen in such droplets that agrees with this but it is not sufficient to drive such motion.  Another possible source of an impulse is that the point of contact with the droplet is weakened at the first point of interface before a ``leaky'' Rayleigh mode attenuates it and this allows surface tension from the back of the drop to pull the droplet in the direction of the sound.  

For a monochromatic wave, every particle follows the same path as every other one with only a phase difference between them.  One can come to many conclusions by halting the perturbation results at different orders for the velocity and density.  By treating them to the same order and correcting the drift at each point, as above, no such complication arises.  The above analysis of Reynolds stress shows that 
attenuation of sound need not impart any momentum.  It merely alters the long range pressure redistribution induced by the sound.  It is interesting that these droplets are often deformed in the direction of motion indicating that the trailing contact point is being pulled up by surface tension.  One possible mechanism for this drift is that the surface acoustic waves passing the leading edge are making contact and pulling it downwards and so may be a result of differential contact and force in the vertical direction.  Another would be in the effects of surface waves on the EZ layer and the separated charge layers that it divides.  A gradient in this layer to a minimum at the point of contact with the surface waves could then create an electric force that drives the droplet in the same direction.  

Using the above results in eqn.~\ref{dP} for the nonlinear pressure correction we see that for a 20cm driver disk in a 30cm cylindrical chamber, with particle velocities of 10cm/s, the pressure difference is $\Delta P\approx10^{7}$Pa$\cdot$s.  For a compressibility of $\alpha\sim10^{-10}$ this gives a density shift $10^{-3}\rho\sim1$mg/cm$^{3}$.  For Poiseuille flow in a tube with a uniform pressure gradient, this gives a source of vorticity at the edge of the beam and a slight deviation from parabolic flow behavior with the support of the beam being slowed not sped up as in acoustic streaming.  Attenuation would give a gradual equilibration of the density down the tube but, without any such external pressure field to drive it, there is no vorticity source at all.  This again supports the notion that the vorticity and flow must arise at the driver itself.  

\subsection{Radiation ``Stress''}
The notion of radiation stress is a problematic concept given that the only true stresses that arise in hydrodynamics are viscous.  It is natural to seek a perturbative approach to the corrections due to the nonlinear advective term but there is not necessarily a unique way to do this.  In the treatment of Stokes drift, we have already introduced a very different approach than the usual ones.  There is no question that the nonlinearity will alter the shape of the acoustic waveforms to low order but the amplitude dependent drift has an effect that seems to disrupt even the linearity of the lowest order analysis.  In sec.~\ref{drift} we considered the kinds of dynamics that can arise when there are no external pressure gradients and forces because we are seeking baroclinic sources of vorticity to ensure a consistent use of such forces for the origin of acoustic streaming.  

\begin{figure}
  \begin{centering}
 \includegraphics[width=3in,trim=10mm 140mm 10mm 40mm,clip]{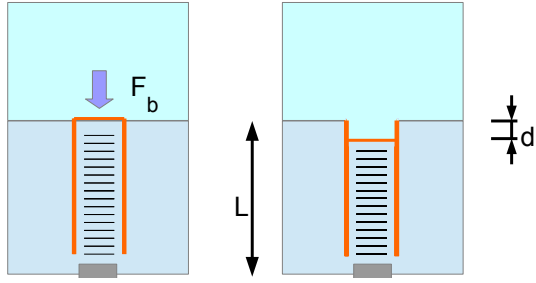}
 \caption{\label{lift} Density change induced surface deformation in an open cylinder.  }
 \end{centering}
 \end{figure}
 
The acoustically induced pressure changes above give us a simpler way to think of these changes than the  usual radiation pressure and stress concepts.  As an example let us consider the case of surface deformation for an acoustic beam at a fluid interface with densities $\rho_{l}$ and $\rho_{u}$ for the upper and lower fluids respectively \cite{Beyer}.  Consider a tube, open at the bottom containing an oscillator as in fig.~\ref{lift}.  Let there be a piston cap at the interface that completely absorbs or reflects the sound wave so none travels into the upper medium.  For a container much wider than the beam, the previous pressure correction of eqn.~\ref{dP} gives a modified density 
\begin{align}
\rho_{l}'=\rho_{l}\left(1+\alpha\frac{a^{2}}{8R}\right)=(1+A)\rho_{l}
\end{align}
and the pressure correction at the interface gives
\begin{align}
\rho_{l}L=\rho_{l}'(L-d)+\rho_{u}d
\end{align}
so that
\begin{align}
\frac{d}{L}=\frac{{A}}{1+A-\frac{\rho_{u}}{\rho_{l}}}\approx\frac{{A}}{1-\frac{\rho_{u}}{\rho_{l}}}.
\end{align}
This provides a simple case to explain the elevation changes at surfaces due to impinging acoustic beams based on a simple equilibration of pressure at the interface.  For more complicated geometries the situation is more difficult as there are parcels of fluid with density variations driven by oscillatory effects that then have to expand and so cost internal energy.  Determining the force on these seems not so trivial.  It is dangerous to try to take energy gradients as a measure of forces.  Certainly forces do arise from energy gradients but using the converse can lead to hopeful and misguided results.  Assuming that there is such a force implies there should be a constant growing circulation around a horizontal beam in a chamber and strong vertical flow about a vertical beam without confining walls as above.  This would generate a continuous attenuation on the beam to drive such a flow.  It seems that a microscopic analysis of the forces on the mass parcels should preclude this but this should receive some more serious analysis as part of a general investigation of acoustic beams oblique on surfaces.

\subsection{Darcy Flow}
For the flow of fluid through topologically complex tortuous structures such as sintered materials, granular packings and extracellular matrices the net flow can often be given by the Darcy law.  In these materials, inertia of fluid motion is irrelevant and the microscopic motion is entirely laminar.  Since microstreaming is dependent on inertia on the scale of the vortices, this eliminates any such effect.  However, there is an impressive body of experimental literature that shows that the flow in these materials can be strongly altered by the presence of ultrasound.  In the case of oil filters this can raise the flow rate by two orders of magnitude after which it exponentially decays to a lower, but usually still higher, rate while the ultrasound is present.  Strangely, this still occurs for the use of clean oil where clogging of the filter should not be an issue \cite{Bjorno}.  We know from bubble rising experiments that very small amounts of impurities can radically change the behavior of flow around a pure air bubble to that around a solid sphere of the same mass.  For this reason, it is unclear if cleanliness is the actual issue with the filter flow and that the ultrasound is simply acting to clean some of the filter paths from impurities or, at least, move them to lower flow pathways.  

In experiments on mouse brains and drug delivery \cite{Lewis}, we can see that ultrasound alone, even without the presence of microbubbles increases the range of drug delivery.  These experiments are done at constant flow rate, not pressure, so it seems that the ultrasound is allowing the fluid to travel along a new set of paths at the expense of others.  Such a system is rather messy and it is possible that the dye is expanding brain vesicles and pushing blood out of the brain case through the vascular network.  In this case, the unassisted migration of drugs is even worse than hoped for.  These experiments demonstrate a fine ``halo'' of enhanced delivery around this region and so ultrasound may be responsible for almost all the actual migration of dye into the extracellular matrix between brain cells.  

The porosity of brain tissue is about 20\% as an upper bound.  This tells us that the size of the channels between cells is on the order of microns.   Water has been shown to exhibit an increase in viscosity associated with wall exclusion effects near some surfaces of over an order of magnitude for some wall materials.  The fluid in the body can be exceptionally dirty in that it can contain very large polymers and charged molecules that can obstruct such small channels.  It is therefore plausible that some microcleaning effect associated with ultrasound and cavitation in these cavities is possible.  We have already seen that at such amplitudes the trough pressures in these sound waves are sufficiently low to allow this to happen.  The time scales of oscillations are very small so, as usual, one has to wonder if this is truly relevant.  

Experiments on the EZ phase of water allow the possibility that it is actually better described as a solid, due to its absence of diffusion, higher density, and ordered structure, so that the apparently higher viscosity from falling particle experiments is actually a kind of fracturing of this order or rearrangement of it about the incident falling particles.  Such a state might completely block motion in small channels.  Recent sleep studies have shown that sleep induces a significant increase in extracellular volume so that waste products can be removed.  This may set some threshold on the size of such an obstruction.  The energy density of ultrasound is very small but the pressure drop may be sufficient to drive cavitation at nucleation sites and destroy such an ordered gel-like phase of water and allow migration through the matrix.  
Direction to probe this problem include controlled cellular matrices with ultrasound at changing orientations and the inclusion of specific centers known to give cavitation at the given pressures and time scales.  An important control for any cavitation result is the external background pressure of the system.  When this becomes large enough cavitation can be halted.  Similarly, lower pressures may enhance the effect.

\section{Afterwords}
The problem of wave flow interactions has quite a few great challenges with technological importance remaining.  The growth in breadth of physics has led to a proliferation and favoring of mathematical shortcuts in the form of perturbation theory and symmetry methods that often overlook microscopic considerations and some rather powerful constraints on consistency one can glean from conservation laws and well posed thought experiments.  Unification has never been more in favor and this can blind us to situations that initially look similar but are fundamentally different.  The behavior of EM waves in media and that of the various waves in hydrodynamics and at the interface with solids have been shown to be such examples.  Many asymmetrical and evanescent and patterned wave analogs to the results of waves in static media remain to be investigated.  

The effects of simple discontinuous shears on sound and EM waves have been discussed but we have left the question of gradual shears and flows largely untouched except for a general basis expansion approach.  The wave-flow decomposition in acoustics has led to many problems involving mass and momentum conservation which are crucial in determining the kinds of forces sound can impart.  A resolution of this to low order has been introduced that eliminates this problem in a far more satisfactory way than the general lagrangian mean but leaves much work to do.    On the microscale the effects of structured water phases are just one of the kinds of ever rich surface physics that must be considered.  Our confidence in this domain has to be blunted by the fact that the larger scale results on acoustic streaming, and even surface wave physics, has never resulted in a clean harmony between theory and experiment.  

Given the growing use of ultrasound as a safe and powerful biological probe and therapy, it has never been more important to understand such basic physical phenomena.  It is this author's opinion that the frequently formulaic and perturbative approach to hydrodynamics is often not successful and, even when it is, gives a poor or misleading level of understanding and that well understood idealized examples will advance the field best at this time.  This may involve some serious reconsideration of past work and revered authors.  Once this is accomplished, I predict we will have not just solid agreement of theory and experiment but theory will regain its place as a powerful and respected predictive tool of phenomenology in the subject.

\section{Acknowledgements}
The author thanks Raghu Ragavan for many stimulating discussions.

\end{document}